\documentclass[apj]{emulateapj}
\usepackage{times}
\usepackage{graphicx}
\usepackage{amssymb,amsmath}
\begin{document}

\title{TTVFast: An efficient and accurate code for transit timing inversion problems}

\author{Katherine M. Deck\altaffilmark{1}, Eric Agol\altaffilmark{2}, Matthew J. Holman\altaffilmark{3}, David Nesvorn\'{y}\altaffilmark{4}}

\altaffiltext{1}{Department of Physics and Kavli Institute for Astrophysics and Space Research,
Massachusetts Institute of Technology, 77 Massachusetts Ave., Cambridge, MA 02139}
\altaffiltext{2}{Department of Physics and Astronomy, University of Washington at Seattle, Seattle, WA}
\altaffiltext{3}{Harvard-Smithsonian Center for Astrophysics, 60 Garden St., Cambridge, MA 02138}
\altaffiltext{4}{Southwest Research Institute, Department of Space Studies,  1050 Walnut St., Suite 400, Boulder, Colorado 80302}

\begin{abstract}
Transit timing variations (TTVs) have proven to be a powerful technique for confirming {\it Kepler} planet candidates, for detecting non-transiting planets, and for constraining the masses and orbital elements of multi-planet systems.  These TTV applications often require the numerical integration of orbits for computation of transit times (as well as impact parameters and durations);  frequently tens of millions to billions of simulations are required when running statistical analyses of the planetary system properties.    We have created a fast code for transit timing computation, TTVFast, which uses a symplectic integrator with a Keplerian interpolator for the calculation of transit times \citep{KOI142}. The speed comes at the expense of accuracy in the calculated times, but the accuracy lost is largely unnecessary, as transit times do not need to be calculated to accuracies significantly smaller than the measurement uncertainties on the times. The time step can be tuned to give sufficient precision for any particular system.  We find a speed-up of at least an order of magnitude relative to dynamical integrations with high precision using a Bulirsch-Stoer integrator.
\end{abstract}
\section{Introduction}\label{sec:Intro}
The number of confirmed planetary systems is growing rapidly, leading to statistical inferences regarding the frequency of planets. However, our precise knowledge of the basic features of individual systems is often still very limited because the major planet detection methods do not individually constrain planetary masses, radii, and orbital configurations.  This has obscured the rich dynamical past of planetary systems and inhibits our understanding of the role various processes play in planet formation and evolution.
%


The most promising method to determine the densities of planets as well as their orbital parameters makes use of the fact that interactions between planets in a multi-planet system produce deviations from Keplerian motion. These interactions are still difficult to detect via radial velocity (though possible, e.g. GJ876, HD 82943 and 55 Cancri \citep{GJ876, GJ8762,Tan,Nelson,Nelson2} ), but because times of transit are often precisely measured, it is possible to detect small transit timing variations, or TTVs, from a constant period Keplerian orbit \citep{HolmanMurray,SteffenAgol}.  These deviations depend sensitively on the masses and orbital configurations of the interacting planets. Since transiting planets have measured relative radii, a relative mass measurement from TTVs can be used to infer constraints on the planetary composition.

TTVs, in combination with transit duration variations (TDVs) or lack thereof, can also yield measurements of the full orbital state of a system from photometry alone (e.g. \citet{Kepler36,KOI142}). Additionally, TTVs have been used to place limits on the presence of companions of Hot Jupiters \citep{Steffen} and to detect and characterize non-transiting planets \citep{Ballard,KOI872, KOI142}. Using TTVs of single-transiting systems to probe the population of non-transiting planets is a promising route towards determining the frequency of non-coplanar systems.  TTVs have also been used to measure the coplanarity of planetary systems with known stellar obliquities \citep{Kep30,Kep56}. 


Measurements of TTVs have become more common with {\it Kepler}: around 3-10\% of the candidate systems show transit timing variations, though not all of these signals will be uniquely invertible \citep{Tsevi,Xie}. The task of modeling these systems is formidable, because inverting a set of transit times, and possibly transit durations, for the masses and orbits of the constituent planets is computationally expensive. At a very basic level, this inversion involves fitting a model of gravitationally interacting planets to the data. However, each evaluation of the model requires integrating an $n$-body system numerically for often hundreds of orbits and determining the times of transit. Even when a best fit solution has been found,  Markov chains used to determine parameter uncertainties can easily require $\gtrsim 10^7$ model evaluations to converge. In cases where parameter space must be searched widely to find a solution, such as when one of the interacting planets is not transiting, the problem is even worse.

 In general, there is no simple analytic solution for a general TTV signal, and so numerical integrations are unavoidable.  In the case of planetary pairs near first order mean motion resonances an approximate formula is known, though there is a degeneracy between masses and free eccentricities \citep{LithwickWu}. Note that this formula cannot be used for single transiting systems with TTVs. The degeneracy can be approached statistically, yielding a measure of the typical eccentricity and mass-radius relationship for pairs of planets near first order resonances, with low free eccentricities \citep{HaddenLithwick}. In principle full numerical integrations or a more accurate formula would break this degeneracy as well,  although to do so the transit times must be measured precisely enough such that higher order effects are observed. 
 
As a result of these difficulties, the information encoded in the TTV signals of the {\it Kepler} candidates regarding planetary masses and orbital parameters has not been fully taken advantage of.  Many of the candidates are not favorable RV follow-up candidates, and there is currently no other viable option for determining the masses and orbits of planets in these systems.  This has motivated us to optimize the basic model evaluation method used by all TTV inversion codes, so that inverting these data sets becomes less computationally demanding. Furthermore, we recognize that many researchers do not have access to an $n$-body integrator which also determines transit times, and so we are also releasing our optimized code so that more scientists can work on inverting interesting TTV signals. Our integration+transit timing code, dubbed TTVFast, is approximately $5-20$ times faster than standard methods (see Section \ref{sec:Numerical}), and is very similar to the code used for modeling the KOI-142 and KOI-872 systems \citep{KOI872,KOI142}. The code can also determine the radial velocity of the star at a set of supplied times for the cases in which both RV data and TTV data is available. Note that this code does not apply to circum-binary planets both because of the splitting for the Hamiltonian chosen (described in Section \ref{sec:inopt}) and because planetary transits of only a single central object are considered.

In Section \ref{sec:Basic}, we discuss the basic structure of our code, and in particular describe how it splits into two parts: the integration of the orbits and the determination that a transit has occurred, and then the transit time calculation and the subsequent determination of the position and velocity of the transiting planet at the time of transit. We discuss how we optimize each part for speed and accuracy. In Section \ref{sec:Numerical},  we show the results for some basic numerical tests, compare our code to a Bulirsch-Stoer integration scheme and more generally consider the robustness of our code.

The code presented here does not determine transit times to as high an accuracy as possible.  We are concerned with balancing computational efficiency with calculating accurate times of transit, and much of our optimization is based on achieving $\sim 1-10$ second precision of transit times.   We discuss how to deal with systems where higher accuracy is needed in Section \ref{sec:Discussion}. \citet{NesvornyMorbidelli}, \citet{N3} and \citet{NB10} have developed an analytic method for inverting TTVs using perturbation theory which we will discuss in Section \ref{sec:analytic}.

Note that in order to be used for inverting a TTV signal, TTVFast needs to be called by whatever minimization routine the user prefers to implement. Publicly available routines for MCMC sampling of a posterior distribution or for estimating the evidence for a particular model could be useful (e.g. emcee: The MCMC Hammer \citep{Foreman} or MultiNest \citep{MultiNest1,MultiNest2}). 

\section{The basic idea}\label{sec:Basic}
Given the dynamical state of the system at a reference time\footnote{The code reads in a set of instantaneous Jacobi orbital elements, astrocentric elements or Cartesian astrocentric positions and velocities.} and planetary masses, relative to the mass of the star, our code integrates the full Newtonian equations of motion for point masses interacting via gravitational interaction. General relativistic and tidal effects are negligible on the time scale of most observations and we do not include them. We use a right-handed coordinate system, where the sky plane is spanned by the orthogonal unit vectors $\hat{x}$ and $\hat{y}$ and is perpendicular to our line of sight. The observer is located at $z = +\infty$, looking towards the origin.

During the integration, the code checks for planetary transits of the host star.  The transit time is defined as when the projected distance on the sky plane between the center of the star and the center of the planet is minimized and the planet is in the foreground (e.g. \citet{JoshWinn}). Therefore, the time of transit satisfies  
\begin{align}\label{dot1} 
D \equiv \vec{r}_{sky} \cdot \vec{v}_{sky} &= x v_x + y v_y = 0,
 \end{align} 
 and $z>0$, where $x,v_x,y, v_y $, and $z$ are astrocentric coordinates of the planet.  This corresponds to the mid-transit time as measured from a transit light curve as halfway between the ingress and egress except when the sky-plane velocity changes significantly along the orbit during the transit. Note that $D$ changes sign even if the planet is not on the face of the star! Therefore the user should check to ensure that $r_{sky}< R_{star}$ at each reported transit. This is important for mutually inclined systems or for planets with grazing orbits.
 
 The integration is performed in Jacobi coordinates, and the quantity  
 \begin{align}\label{dot} 
 D' &= \vec{r}'_{sky} \cdot \vec{v}'_{sky} = x' v'_x + y' v'_y, 
 \end{align}
  is tracked, where primes denote Jacobi coordinates. The difference between Jacobi and astrocentric coordinates is of order $m_{planet}/m_{star}$ (e.g. \citet{WisdomHolman}) and is in general not important for determining whether a transit occurs; see Section \ref{sec:TTDet} for a discussion). 
  
  If during a time step $D'$ changes sign from negative to positive and $z'>0$, a transit - according to the definition given in Equation \eqref{dot1} - has occurred. The user should be aware that $D'$ changes from negative to positive at occultation as well, but since $z'<0$ at this time occultations cannot be confused with a transit. The fact that $D'$ changes from negative to positive at transit and at occultation implies that there must be two other roots present in the orbit - in between transit and occultation, and then in between occultation and transit, where $D'$ changes from positive to negative.  Since the code only looks for sign changes from negative to positive, these points will not be confused as transits either. However, the time step must be small enough such that the transit condition can be met. If, for example, the time step was large enough to include two subsequent roots, $D'$ would not change sign even though a transit occurred during that step. In practice this only occurs for very eccentric orbits using large time steps, and is avoided by using a smaller time step. Given that interacting planets with eccentric orbits require smaller time steps in general (see Section \ref{sec:inopt}), we don't believe this problem will arise often in practice. 
  
  When $D'$ changes sign from negative to positive and $z'>0$,  the transit time is then determined, as well as the orbital state (sky-projected astrocentric position and velocity) of the planet at the time of transit. These quantities can then be used to determine the duration of the transit (making the constant velocity approximation during the transit) or be used as input to a photometry model (if the user is directly fitting a light curve and not the intermediate quantities of transit time and duration, so-called ``photo-dynamics").  
 
 This scheme can be optimized in two parts. First, the numerical integration of the orbits must be performed for the time span of the observations. Second, once per orbital period of a transiting planet, the code must efficiently and accurately determine the time of transit and the orbital state of that planet at transit.  These processes are linked in that the transit finding cannot proceed without the output of the integration, but they are decoupled in that they can be considered separately for optimization.

\subsection{Integration method and optimization}\label{sec:inopt}
The specific integration algorithm and the order to which it is carried both affect the efficiency and accuracy of the integration. Nearly integrable Hamiltonian systems can be evolved more efficiently using a symplectic mapping compared with standard integration schemes like Bulirsch-Stoer \citep{WisdomHolman,Mclachlan}. An $n$-planet system evolving through gravitational interaction falls into this category, as the motion of the planets is nearly a sum of $n$-Keplerian ellipses. 

 The full Hamiltonian for the $n$-body problem ($n$-1 planets) can be written exactly as
\begin{align}\label{Hamiltonian}
H & = \sum_{i=1}^{n-1} H_{Kepler,i} +H_{interaction} \nonumber \\
H_{Kepler,i} \equiv H_A & = \frac{p^{'2}_i}{2 m'_i} - \frac{G m_i m_0}{r'_i} \nonumber \\
H_{interaction} \equiv \epsilon H_B & = \sum_{i=1}^{n-1} G m_i m_0\bigg( \frac{1}{r'_i} -\frac{1}{r_{i0}}\bigg)  \nonumber \\ 
&- \sum_{0<i<j} \frac{G m_i m_j}{r_{ij}}
\end{align}
where again primes denote Jacobi coordinates and momenta, $m'$ denotes a Jacobi mass, $m_0$ is the stellar mass, and $r_{ij}$ denotes the Euclidean distance between bodies $i$ and $j$ (see e.g. \citet{WisdomHolman} for the definition of Jacobi masses and coordinates; another splitting must be used for circum-binary planets). Note that $H_{interaction}$ is of order $\epsilon \sim {\rm max}_{i>0}\{m_i/m_0\}$ smaller than $H_{Kepler}$. We will denote $H_{Kepler} \equiv H_A$ and $H_{interaction} \equiv \epsilon H_B$. 

The value of a phase space function $f(Q,P)$, where $(Q,P)$ are the phase space variables,  after evolution for  a time $\Delta t$ according to the Hamiltonian $H(Q,P)$ can be written as 
\begin{align}
&f(t_0+\Delta t)=  e^{L_H\Delta t} f \bigg\lvert_{t=t_0} \nonumber \\
 &\approx f\bigg\lvert_{t=t_0} + \Delta t \{ f,H\}\bigg\lvert_{t=t_0} + \frac{\Delta t^2}{2} \{\{f,H\},H\}\bigg\lvert_{t=t_0} + \ldots 
\end{align}
%
where $L_H$ denotes the evolution operator for the corresponding Hamiltonian $H(Q,P)$ (see, for example, Hairer (2006)), and $\{\ldots, \ldots \}$ denotes a Poisson Bracket. In our case, the Hamiltonian $H$ is given as $H = H_A +\epsilon H_B$.

 The entire Hamiltonian does not have an analytic solution, but each piece of the Hamiltonian, $H_A$ and $H_B$, are exactly solvable independently. The Wisdom-Holman mapping makes use of the fact that both $H_A$ and $H_B$ lead to motion which is efficiently computable.  The problem is split into impulsive kicks (due to the planetary interactions) interleaved with Kepler steps, in many cases allowing the integrator to take as few as $\sim 20$ steps per orbit of the innermost planet. 
 
 We use a leap frog approximation (denoting $L_{H_A} = A$ and $L_{H_B} = B$)
   \begin{align}\label{leapfrog}
e^{L_H \Delta t} = e^{(A+\epsilon B) \Delta t} & \approx e^{A\Delta t/2} e^{ \epsilon B \Delta t }e^{A \Delta t/2} 
 \end{align}
Since $e^{A \Delta t}$ is the operator which evolves a phase state function according to $H_A$ for a time $\Delta t$ - according to purely Keplerian motion - and since $e^{\epsilon B \Delta t}$ is the operator which evolves a phase state function according to interaction Hamiltonian $\epsilon H_B$ for a time $\Delta t$, the leap frog operator given in Equation \eqref{leapfrog} exactly translates to evolving the system according to only the Keplerian Hamiltonian, $H_A$ for half a time step, followed by evolution according to the interaction Hamiltonian $\epsilon H_B$ for a full time step, and then another half step of the Keplerian evolution.

The Keplerian step is carried out using the Gauss f and g functions and will be explained further in Section \ref{sec:kep}. Care must be taken when evaluating numerically the difference between the like quantities $r_i'^{ -1}$ and $r_{0i}^{-1}$ in $H_B$. Since the interaction Hamiltonian is independent of the momenta, evolution according to it only alters the velocities and does not change the positions of the bodies - this is why it is referred to as an impulsive kick step.

 The leap frog scheme has a dominant error term of order $\epsilon \Delta t^3/24 \{A,\{A,B\}\}$, the other error term cubic in $\Delta t$ is $\epsilon$ smaller: $\epsilon^2 \Delta t^3/12 \{B,\{A,B\}\}$ (and the next largest error terms are of order $\Delta t^5$). The other leap frog scheme (interaction kick, Keplerian drift, interaction kick) is equally valid but has an error term twice as large ($\propto \epsilon \Delta t^3/12$). 
 
 
 Two consecutive leap frog steps can be combined as
    \begin{align}\label{leapfrogCombo}
 e^{(A+\epsilon B) 2 \Delta t} & \approx e^{A\Delta t/2} e^{ \epsilon B \Delta t }e^{A \Delta t} e^{ \epsilon B \Delta t }e^{A \Delta t/2} 
 \end{align}
 as long as precise output is only needed at the endpoint. This reduces the number of applications of the Keplerian drift operator (though it slightly complicates the code when a transit is detected - see Section \ref{sec:Bracket}), so that both the kick and drift are applied approximately an equal number of times.

 Higher order integrators (involving more than three total applications of $e^{\Delta t a A}$ or $e^{\Delta t b\epsilon B}$, where $a$ and $b$ are constants) increase the accuracy of the integration (e.g. \citet{LaskarRobutel}). However, the transit times found using the leap frog method (with use of a corrector, see below) have a dominant error term that is not from the integration itself, but typically from the method used to determine a transit time once in the vicinity of a transit, as explained in Section \ref{sec:scale}. Therefore, the increased accuracy obtained with higher order integrators will not improve the overall accuracy of the transit times in general.
 
 We recommend that a maximum time step of $\Delta t= P_{inner}/20$ should be used for the symplectic integrator. Larger steps can lead to step-size chaos and inaccurate orbits \citep{WisdomMatt}. Additionally, for eccentric orbits, a smaller step is required than for nearly circular orbits in order to resolve pericenter passage. \citet{RauchHolman} suggest a time step no larger than $1/20$ the orbital period the planet would have if it orbited at a constant semimajor axis equal to the pericenter distance $a(1-e)$. 
 
 \subsubsection{Symplectic Correctors}
 The leap frog scheme can be shown to exactly correspond to evolving the equations of motion derived from a mapping Hamiltonian $H_m$, that is
 \begin{align}
 e^{\Delta t L_{H_m}} & = e^{A\Delta t/2} e^{ \epsilon B \Delta t }e^{A \Delta t/2}.
 \end{align}
 This operator exactly evolves the equation of motion of the Hamiltonian 
 \begin{align}\label{mapping}
H_m & = H_{A}+ \frac{2 \pi}{\Omega} \epsilon H_{B}   \sum_{k=-\infty}^{\infty} \delta(t-\frac{2\pi k}{\Omega}) 
\end{align}
where $\Omega = \frac{2\pi}{\Delta t}$ and $\Delta t$ is the time step of the mapping. The Fourier series of the comb of Dirac delta functions is an infinite sum of equally weighted sines and cosines with frequency $k \Omega$, where $k$ is any integer including zero,  and hence the mapping Hamiltonian only differs from the true Hamiltonian by high frequency terms. The largest physical frequency of the planetary problem is approximately the largest planetary mean motion. Therefore the higher frequency terms contributed by the delta functions - which are the difference between the mapping Hamiltonian and the real one - will average out on timescales comparable to the orbital period and longer. 

\citet{WisdomField} realized that there was a canonical transformation between the real Hamiltonian and the mapping Hamiltonian. This canonical transformation between ``real" and ``mapping" phase space variables, when applied, reduces the error of the integration method in approximating the motion of the real system. The corrected leap frog step takes the form
\begin{align}
 e^{\Delta t L_{H}} &\approx  e^{-C} e^{A\Delta t/2} e^{ \epsilon B \Delta t }e^{A \Delta t/2} e^{C} \bigg( = e^{-C}  e^{\Delta t L_{H_m}} e^{C} \bigg)
\end{align}
so that the phase space state in real coordinates is transformed to mapping coordinates (by $e^{C}$), evolved according to the mapping Hamiltonian, and then transformed back to real coordinates (by $e^{-C}$ - the inverse corrector). These correctors are intuitively considered canonical transformations, but more literally they are chosen to be a combination of interaction kicks and Keplerian drifts such that the higher order error terms of the leap frog scheme are canceled out. These correctors are costly, as even the lowest order corrector involves twelve Kepler steps or interaction kicks.

We find that applying the corrector only once at the beginning of the integration significantly improves the accuracy of the calculated times. The reason for this is that the mean orbital period of the transiting planets in mapping coordinates is not the same as in real coordinates - and the subsequent error in the transit time accumulates linearly over the entire integration.  The corrector slightly modifies the initial conditions, and therefore slightly modifies the mean orbital period of the transiting planet, reducing this error. When a transit occurs, a copy of the dynamical state of the system is evolved for a small fraction of a time step to determine the transit time, and hence the difference between mapping and real coordinates does not accumulate and the inverse corrector is unnecessary. As a result, we do not use the inverse corrector at every transit (and effectively find the transit time of the mapping system).

We have implemented the third order corrector given in \citet{Wisdom2006}, which removes the dominant error term $\epsilon \Delta t^3/24 \{A,\{A,B\}\}$.  Even the third order corrector does result in very small differences between the mean orbital periods of the planets in real and mapping coordinates. In principle, higher order correctors can be used to remove even more of the linear trend in the error in transit times, with a negligible increase in computational time (since the corrector is only applied once); however, these higher order corrections are only useful if the difference in initial conditions is the dominant source of error even after the third order corrector is applied. Again, this is not the case: the error is in general dominated by approximations made in calculating the transit times themselves (see Section \ref{sec:scale}). 



\subsubsection{Optimizing solving Kepler's equation}\label{sec:kep}
The Keplerian motion of a planet is evolved using the Gauss $f$ and $g$ functions (e.g. \citet{Danby}). This requires determining, for a given a time step $\Delta t$ and the corresponding change in the mean anomaly of a planet $\Delta M = \frac{2 \pi}{P} \Delta t$, what the change in the eccentric anomaly $\Delta E$ is. The answer is the root of the incremental Kepler's equation, given by
\begin{align}\label{IKE}
F(\Delta E) & = \Delta E + 2 \sin{^2\bigg( \frac{\Delta E}{2}\bigg)} e \sin{E} \nonumber \\
&- e\sin{\Delta E}\cos{E} - \Delta M =0
\end{align}
where $E$ is the value of the eccentric anomaly at the start of the time step. 
Each application of the Kepler step requires solving a version of the incremental Kepler's equation \eqref{IKE} $n$ times (for an $n$ planet system).  This is the rate-determining part of the Kepler step. We follow \citet{Danby} and use the first, second, and third derivatives in an extension of Newton's method for root finding (which we can call Danby's method). Danby's method yields quartic convergence - the number of correct digits quadruples after each iteration. Our approach is to use the solutions for $\Delta E$ from the previous three Keplerian steps in a quadratic extrapolation to make an initial guess for  the next value of $\Delta E$. This initial guess is typically correct to 3 or 4 digits already, and hence one iteration of Danby's method yields close to machine precision for the solution. 


\subsubsection{Bracketing the transit}\label{sec:Bracket}
As discussed above, we reduce the number of applications of the Kepler step by combining subsequent half steps together (as in Equation \eqref{leapfrogCombo}). Before each application of the operator $e^{A \Delta t} e^{B \Delta t}$, we save a copy of the full dynamical state (Jacobi positions and velocities of all the planets) as $p_{behind}$. After the application, we check if a transit has occurred (using the quantity \eqref{dot}). If it has, we copy the updated state to $p_{ahead}$. We wish to have knowledge of the dynamical state of the system on either side of the transit. However, the error in the states $p_{behind}$ and $p_{ahead}$ is large ($\sim \epsilon \Delta t^2$) because the output is not at the conclusion of a symplectic time step, so we must apply the operator $e^{-A \Delta t/2}$ to each state, for example,
\begin{align}
\tilde{p}_{behind}&= e^{-A \Delta t/2} p_{behind}
\end{align}

The states $\tilde{p}_{behind}$ and $\tilde{p}_{ahead}$ now have error of order ($\sim \epsilon \Delta t^3$). However, since the states have now been integrated backwards along a Keplerian arc for half a time step, the transit may no longer be bracketed. We check the quantity $D'$ given in \eqref{dot} for the transiting planet using $\tilde{p}_{behind}$ and $\tilde{p}_{ahead}$. If $D'$ has changed sign, the transit is still bracketed, and $\tilde{p}_{behind}$ and $\tilde{p}_{ahead}$ are accurate versions of the dynamical state on either side.  If the transit is no longer bracketed, we set $\tilde{p}_{behind}(\mbox{new}) =  \tilde{p}_{ahead} = e^{-A \Delta t/2} p_{ahead}$, and step $p_{ahead}$ forward to complete a full symplectic time step so that the new $\tilde{p}_{ahead}$ is given by
\begin{align}
\tilde{p}_{ahead} (\mbox{new})&= e^{A \Delta t/2}  e^{B \Delta t}  p_{ahead} \nonumber \\ 
\end{align}
Again, then, we will have the transit bracket accurately by $\tilde{p}_{behind}$ and $\tilde{p}_{ahead}$.


Because of the extra Kepler steps required at a transit, slightly more applications of the Kepler step are required compared to the kick step. The Kepler step requires a comparable amount of computational time compared to the interaction kick step for a typical number of planets  ($n \lesssim$ 5-10). 

\subsection{Transit time, impact parameter, and transit duration determination}\label{sec:TTDet}

Given the dynamical state of an $n$-planet system at two times bracketing a transit (separated by $\Delta t$), what is the most efficient way to determine the time of transit of one of the planets? When using a standard integration scheme, transit times are often determined by solving for the time at which $D(t) =0$, or when the projected distance between the star and the planet is at a minimum. This can be achieved using any root finding algorithm, but requires an integration of the equations of motion for amounts of time $\Delta < \Delta t$ to the time $t$ whenever the function $D(t)$ is evaluated.

Recall that the symplectic integration exactly solves the equations of motion of the Hamiltonian $H_m$, given in Equation \eqref{mapping}. The Hamiltonian $H_m$ depends on the time step $\Delta t$, and therefore changing the time step involves changing the system being solved. In order to accurately change the time step mid-integration, one would need to convert from mapping coordinates for a particular $\Delta t$ back to real coordinates, and than back to mapping coordinates for the new $\Delta t$ (see e.g. \citet{Kaib}). 

We find it significantly faster to make an approximation which allows us to avoid directly integrating the orbits to zero in on the transit time.  Since the transit time is bracketed within one time step ($\Delta t \le P_{inner}/20$), the motion can be treated as Keplerian during and just around the transit.  This approximation is even more accurate for the outer planets. The following scheme is performed with copies of the dynamical state, so that there is no accumulated error.

First, the state $\tilde{p}_{behind}$, just before the transit, is converted to astrocentric coordinates. Then, we solve for the time of transit, approximating the motion of the transiting planet as Keplerian, and making use again of the $f$ and $g$ functions as follows.
The $f$ and $g$ functions evolve the state of the transiting planet as
\begin{align}
X & = f(\Delta E) x + g(\Delta E) v_x \nonumber \\
Y & = f(\Delta E) y + g(\Delta E) v_y \nonumber \\
V_X & = \dot{f}(\Delta E) x + \dot{g}(\Delta E) v_x \nonumber \\
V_Y & = \dot{f}(\Delta E) y + \dot{g}(\Delta E) v_y
\end{align}
where capital letters denote the updated state, and lowercase letters denote the state just before the transit. Therefore, the function we are seeking the root of, $D$ from equation \eqref{dot1}, can be written as
\begin{align}\label{esoln}
D &= XV_X +Y V_Y \nonumber \\
&= f \dot{f} r_{sky}^2 + g \dot{g} v_{sky}^2 + (g \dot{f} +f \dot{g}) (x v_x +y v_y)
\end{align}
The value for $\Delta E$ such that $D(\Delta E)=0$ is solved for using Newton's method. The number of iterations could be reduced using a higher order method (like Danby's, which we use for the Kepler solver), but in practice this transit time method takes only a small fraction of the total computational time. 
The value of $\Delta E$ which solves this equation is related to $\Delta M$ by the incremental Kepler's equation as 
\begin{align}
\Delta M & = \Delta E + 2 \sin{^2\bigg( \frac{\Delta E}{2}\bigg)} e \sin{E_0}-\sin{( \Delta E)} e \cos{E_0} 
\end{align}
where $e$ and $E_0$ are the eccentricity and  the eccentric anomaly at the original state. The change in time from the bracketing point to the transit is then simply $\Delta M$ divided by the mean motion. In this process, we use a Kepler constant of $G (m_\star + m_p)$.   

Note that once we have calculated the change in the eccentric anomaly $\Delta E$ between the bracketing time step and the transit time predicted by a Keplerian orbit - the solution of Equation \eqref{esoln} - we can also find the full dynamical state of the transiting planet at the transit time using the Gauss $f$ and $g$ functions. From this, we calculated the sky projected astrocentric distance $r_{sky}$ and the sky projected astrocentric velocity $v_{sky}$. 

We proceed the same way from the bracketing state ahead of transit, and linearly weight the two predictions for the transit time. The weight of $p_{behind}$ is 1 if the transit occurs at the initial time, and 0 for $p_{ahead}$, and vice-versa in the opposite case. More explicitly, the transit time $\tau$ is:
\begin{equation}
\tau = \frac{(\tau_{behind}-t)\tau_{ahead} + (t+\Delta t-\tau_{ahead})\tau_{behind}}{(t+\Delta t-\tau_{ahead})+(\tau_{behind}-t)}
\end{equation}
where $t$ is the time at the initial bracketing point (corresponding to $p_{behind}$), $t+\Delta t$ is the time at the final bracketing point, $\tau_{behind}$ is the transit time as predicted by the Keplerian arc corresponding to $p_{behind}$, and similarly for $\tau_{ahead}$. 

Note that this interpolation scheme was originally developed by \citet{KOI142} for the TTV analysis of KOI-142b.

In the exact same manner, we weight the values for $r_{sky}$ and $v_{sky}$ determined based on the Keplerian approximation from the two bracketing points. The resulting value for $r_{sky}$ can be used to calculate the impact parameter of the planet, and in combination with $v_{sky}$ can be used to estimate the transit duration. 

Finally, we note that there are some cases with very large eccentricities when this algorithm does not find the correct root of $D$ in Equation \eqref{dot1}. Recall that there are four roots of $D$ per orbit, and Newton's method does not constrain the root to lie within a certain interval. After a root is found using Newton's method, we check that the derivative of $D$ is positive and that $z>0$. If this is not satisfied - if the incorrect root has been found - the code recalculates the transit time using the bisection method. The convergence of the bisection method is slower compared with Newton's method, but it has the advantage that the root found is guaranteed to lie within the bracketing interval.

Lastly, we point out that if the transit falls very close to the end or start of a time step, the evaluated quantity $D'$ can be small enough such that the difference between Jacobi and astrocentric coordinates, or the difference between $D'$ and $D$, is of the same order as $D$ itself. In this case, our condition for a transit, which is calculated in Jacobi coordinates, may be met even though the transit itself does not lie in that time interval. In this case, Newton's method will generally find the correct root regardless, and the correct transit time is returned. However, if this case coincides with the high eccentricity case mentioned above, when Newton's method is may not find the correct root, the transit may not lie within the window passed to the bisection root solver.  In this case, the code returns a default error value. This appears to be such an unlikely situation that we do not alter the code to account for it.

\subsection{Theoretical scaling of the error in transit times}\label{sec:scale}
In total, there are three sources of error in the calculated transit time (ignoring numerical round-off). These are 1) the error in the initial conditions, 2) the error in the state of the system bracketing the transit resulting from the integration itself, and 3) the error resulting from the Keplerian approximation in finding the times of transit. With the corrector implemented, the error in the initial conditions is of order $\epsilon \Delta t^5$ and $\epsilon^2 \Delta t^3$. This error in the initial conditions leads to an error of the same magnitude in the mean motion, which in turn leads to accumulating errors in the transit times of  order $\epsilon \Delta t^4$ and $\epsilon^2 \Delta t^2$. If no corrector were used, the error would be dominated by the $\epsilon \Delta t^2$ term. 

For short periods of time, the true mean motion of the planet can be written roughly as 
\begin{equation}
n(t) \sim n_0[1+\mathcal{O}(\epsilon) t/P_0],
\end{equation}
where $n_0$ is the instantaneous mean motion at a reference time (taken to be one of the bracketing times) and $P_0 = 2\pi/n_0$. In the Keplerian approximation, $\epsilon=0$. The term $\mathcal{O}(\epsilon) t/P_0$ indicates that the difference between a Keplerian orbit with $n(t) = n_0$ and the true orbit grows approximately linearly in time over short enough intervals.  In the simplest case, the motion is circular, and so the change in the true anomaly $f$ of the transit planet is equal to the change in the mean anomaly $\theta$. From the bracketing point, the planet must sweep out a change $\delta \theta$ to reach mid-transit, where 
\begin{equation}
\delta \theta  =\int n(t) dt \sim n_0[t+\mathcal{O}(\epsilon) (t/P)(t/2)],
\end{equation}
 while in the Keplerian approximation the angle is incorrectly estimated to be $\delta \theta_{Kepler} = n_0 t$.   Therefore the error in the resulting transit time $\delta t$ will be 
 \begin{align}
 \delta t \sim \delta \theta/n_0 &\sim n_0\mathcal{O}(\epsilon) (t/P)(t/2)/ n_0 \nonumber \\
 &\propto \epsilon t^2/P \propto \epsilon \Delta t^2,
 \end{align} since $t$ is of order $\Delta t$. This scaling with the time step was verified numerically. Using two bracketing points keeps the same scaling of the transit time error (as $\Delta t^2$) but increases the accuracy of the estimate (by decreasing the coefficient, such that the error is a small coefficient$\times \epsilon \Delta t^2$).  

At the two bracketing times, the state itself has a dominant error of order $\epsilon \Delta t^3$ (since no inverse corrector is applied), and hence the value $n$ used for $n_0$ is incorrect by an amount $\epsilon \Delta t^3$. This yields additional errors of $\epsilon \Delta t^4$ in the transit time.

Therefore, we predict that the transit timing error should typically scale as $\Delta t^2$. This explains why the inverse corrector is not necessary at each transit, while the initial application is, and why higher order correctors and integrators are not helpful. Depending on $\epsilon$, $\Delta t$, and the coefficient (of the error term from the Keplerian approximation) there may be regimes where the other sources of error are more important. 

 \subsection{Radial velocity measurements}
If the CALCULATE\_RV flag is set, the code expects a set of times at which to determine radial velocities for the star. During the integration, if a time of RV observation $t_{RV}$ is passed during a time step, the code evolves a copy of the state to the time of RV observation approximating all orbits as Keplerian (by using the operator $e^{A( t_{RV} - t_0)}$, where $t_0$ is the time at either the beginning or end of the time step, depending on which is closer to $t_{RV}$).  We do not convert between mapping coordinates and real coordinates, and we do not use any correctors to minimize the error incurred by changing the time step. When the state has been evolved to the correct time, the barycentric radial velocity of the star is returned. Please note that the radial velocity is returned as $-v_{z}$ to keep with the convention that the RV is positive when the star is moving away from us. 
\section{Numerical Tests}\label{sec:Numerical}

\subsection{Convergence with Time Step}
We performed basic tests to confirm how the timing accuracy improved as the time step decreased, for a range of $\epsilon$, where $\epsilon$ depends on the masses of the planets relative to the mass of the star and the distance between the planets.  The parameter $\epsilon$ should scale monotonically with the relative TTV (the full amplitude of the TTV relative to the orbital period), though other factors, such as how close a system is to resonance, also affect the TTV amplitude without affecting $\epsilon$.

\begin{figure}[ht]
\begin{center}
 \includegraphics[scale=0.35]{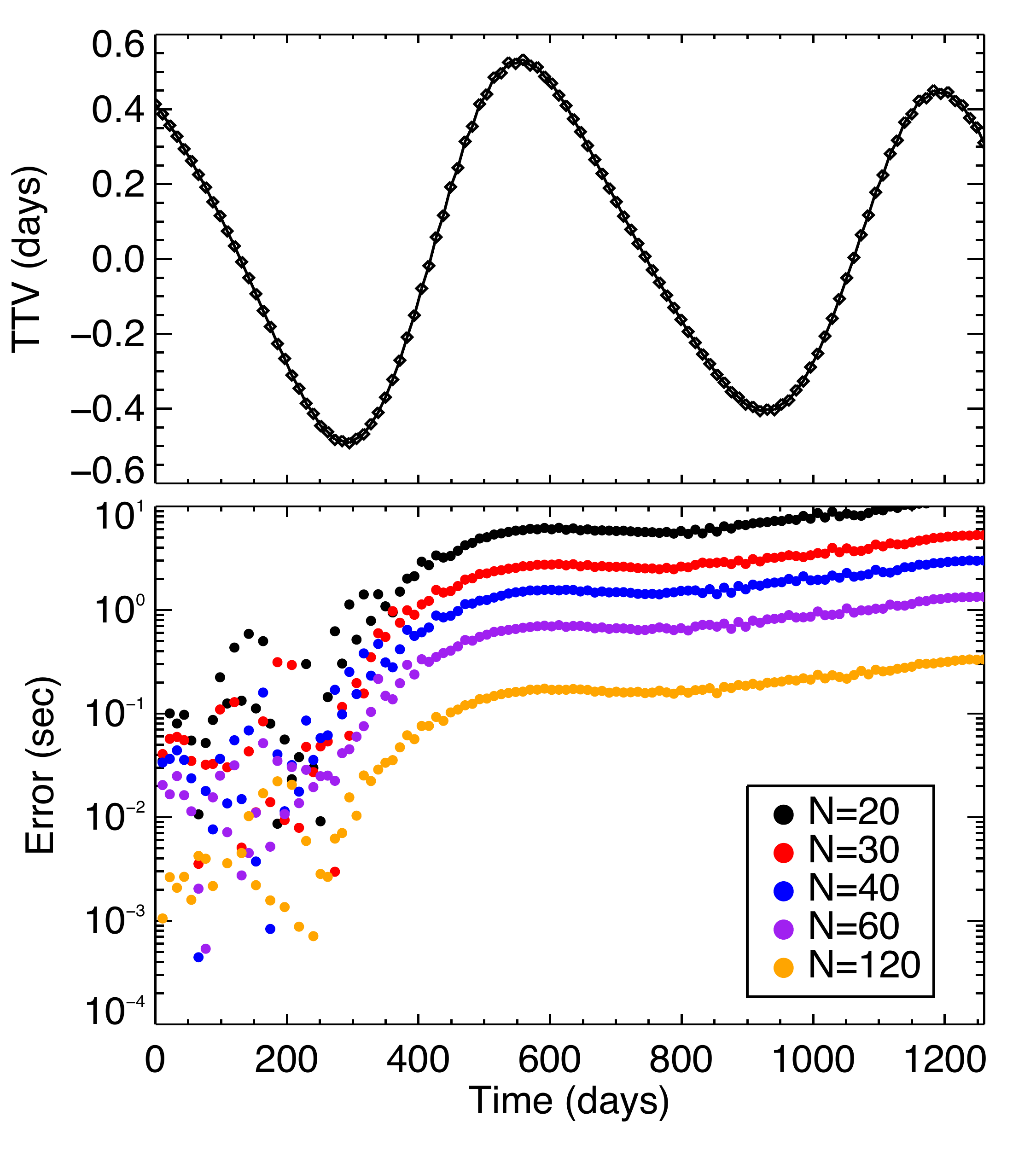}
\caption{Application of TTVFast to KOI-142. In the upper panel, the TTVs of the best fit solution for KOI-142 found by \citet{KOI142} are shown. These are the largest amplitude perturbations measured in the {\it Kepler} sample. Below we show the error, in seconds, between TTVFast and Bulirsch-Stoer. The small amplitude, short period (``chopping") signal of several minutes present in the times of KOI-142b are well resolved. The different colors corresponds to the number of steps taken per orbit by the TTVFast integrator. Note that the error changes by a factor of four when the number of steps doubles, as predicted in Section \ref{sec:scale}. }
\label{fig:KOI142}
\end{center}
\end{figure}
First, for reference, we show in the upper panel of Figure \ref{fig:KOI142} the TTVs of the best fit solution for KOI-142 found by \citet{KOI142} based on the analysis of {\it Kepler} data from Q0-Q14. KOI-142 has one of the largest value of the TTV amplitude observed, nearly 10\%, and a value of $\epsilon \sim 6\times10^{-4}$ \citep{KOI142}. We also show the error in the transit times determined by TTVFast for a different number of steps per orbit $N$ ($\Delta t = P_{KOI-142b}/N$, where KOI-142b is the inner planet). As few as 20 steps per orbit still results in transit times accurate to within 10 seconds. 
\begin{figure}[ht]
\begin{center}
 \includegraphics[scale=0.35]{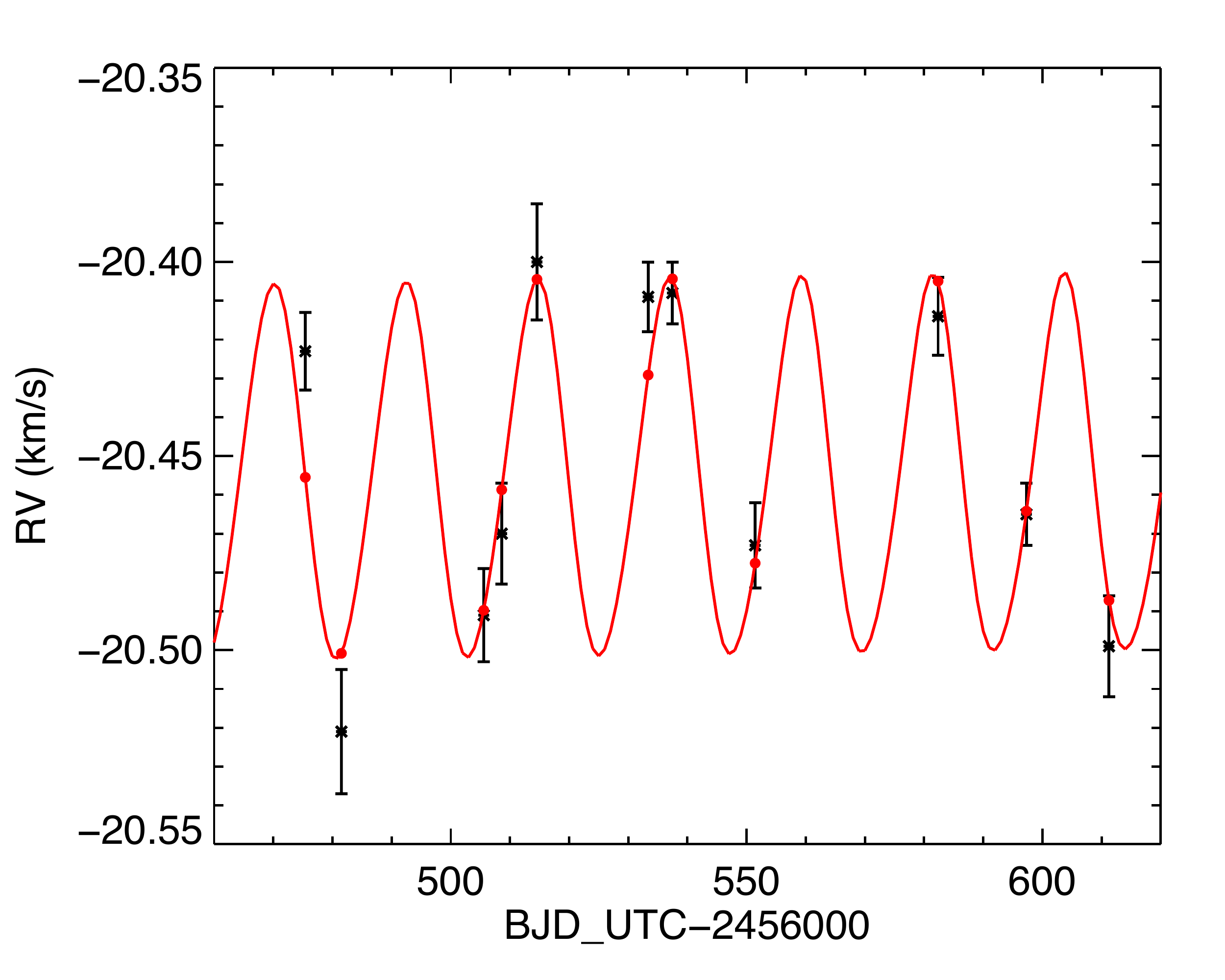}
\caption{Predicted and observed radial velocities for KOI-142. In black are the data, as reported by \citet{Barros}. The red points denote the radial velocities predicted by the best-fit solution for KOI-142 (also used to generate Figure \ref{fig:KOI142}) at the times of the RV measurements using TTVFast.  The underlying red curve shows the continuous radial velocity signal. Note that we used the RV offset reported by \citet{Barros}.  }
\label{fig:RV}
\end{center}
\end{figure}

In Figure \ref{fig:RV}, we show the results of TTVFast for the corresponding radial velocity predictions for KOI-142, employing the same initial conditions used to generate Figure \ref{fig:KOI142}.  The observed radial velocities reported in \citet{Barros} are shown as well. \citet{Barros} found that the observed radial velocities agreed with the amplitude and period predicted by \citet{KOI142}. Figure \ref{fig:RV} makes it clear that the phase of the best fit TTV solution is an excellent match to the RV observations as well.

We next test how the code behaves more generally. Our sample system in this case consisted of two nearly coplanar planets orbiting a solar mass star with orbital periods of 15 and 31.77 days and eccentricities of 0.02. The masses were varied from sub Earth mass to $\sim 700$ Earth masses to vary $\epsilon$, while keeping the mass ratio between the planets fixed. All integrations were performed for $\sim 3,000$ days or approximately 200 orbits of the innermost planet.

\begin{figure}[ht]
\begin{center}
 \includegraphics[scale=0.4]{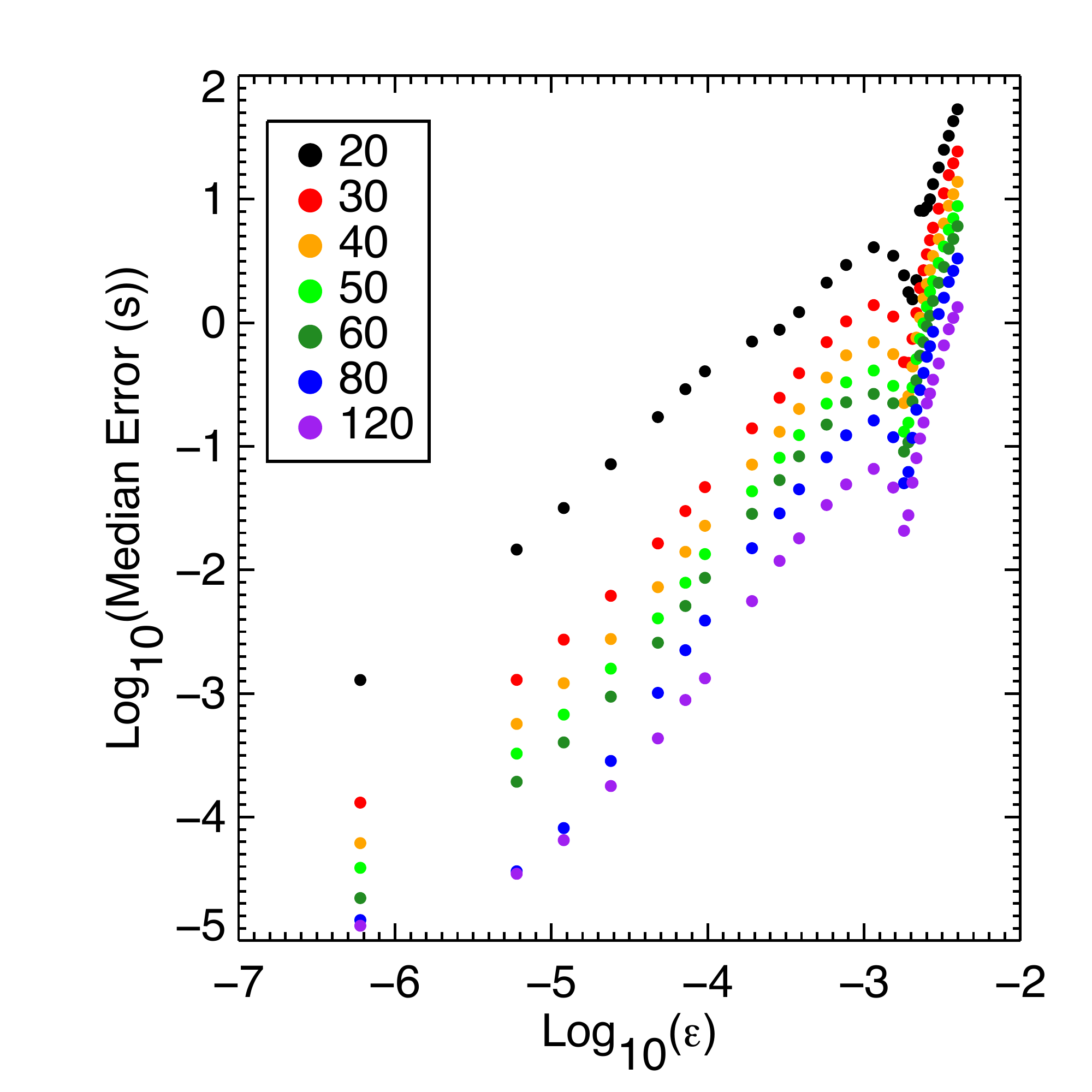}
\caption{Median errors in the transit times of the inner planet after $\sim 200$ orbits, as a function of the the mass of the more massive perturber relative to the mass of the star ($\epsilon$) for varying number of steps per orbit (the different colored dots).  Across a wide range of $\epsilon$, the error scales approximately quadratically in $\Delta t = P_{inner}/N$.  }
\label{fig:scaling}
\end{center}
\end{figure}

In Figure \ref{fig:scaling}, we show the error in the transit times of the inner planet as a function of $\epsilon$ for a different number of steps per orbit $N$.  The errors in the transit times of the outer planet will be of a similar magnitude. This indicates that for Jupiter mass planets in wider orbits, 20 steps per orbit should achieve 10s accuracy.  Closer pairs of planets would likely require more steps per orbit since $\epsilon$ also depends on the inverse of the distance between the planets. This plot confirms that a KOI-142 like system requires only $\sim$ 20 steps per orbit for 10s accuracy, as found in Figure \ref{fig:KOI142}. 

Overall, across a range of $\epsilon$, we find that the scaling of the errors agrees well with $\Delta t^2$. For $\epsilon \lesssim 10^{-3}$, the error scales approximately linearly with $\epsilon$, though at larger values the scaling is much steeper than we predicted in Section \ref{sec:scale}. When exploring even larger masses, many of the test systems were unstable on short timescales. 

Changing the orbital periods of the planets, while preserving the masses of the planets and initial orientation of the orbits, scales the entire system in a self similar way. In this case, $\epsilon$ remains the same, but the error in the transit times grows linearly with the period of the planet. This implies that systems with an inner planet on a longer period orbit will require more steps per orbit to reach a specific timing accuracy than that same system scaled down to shorter orbital periods. 
\subsection{Comparison to Bulirsch-Stoer}\label{sec:BS}

\begin{figure}[ht]
\begin{center}
 \includegraphics[scale=0.4]{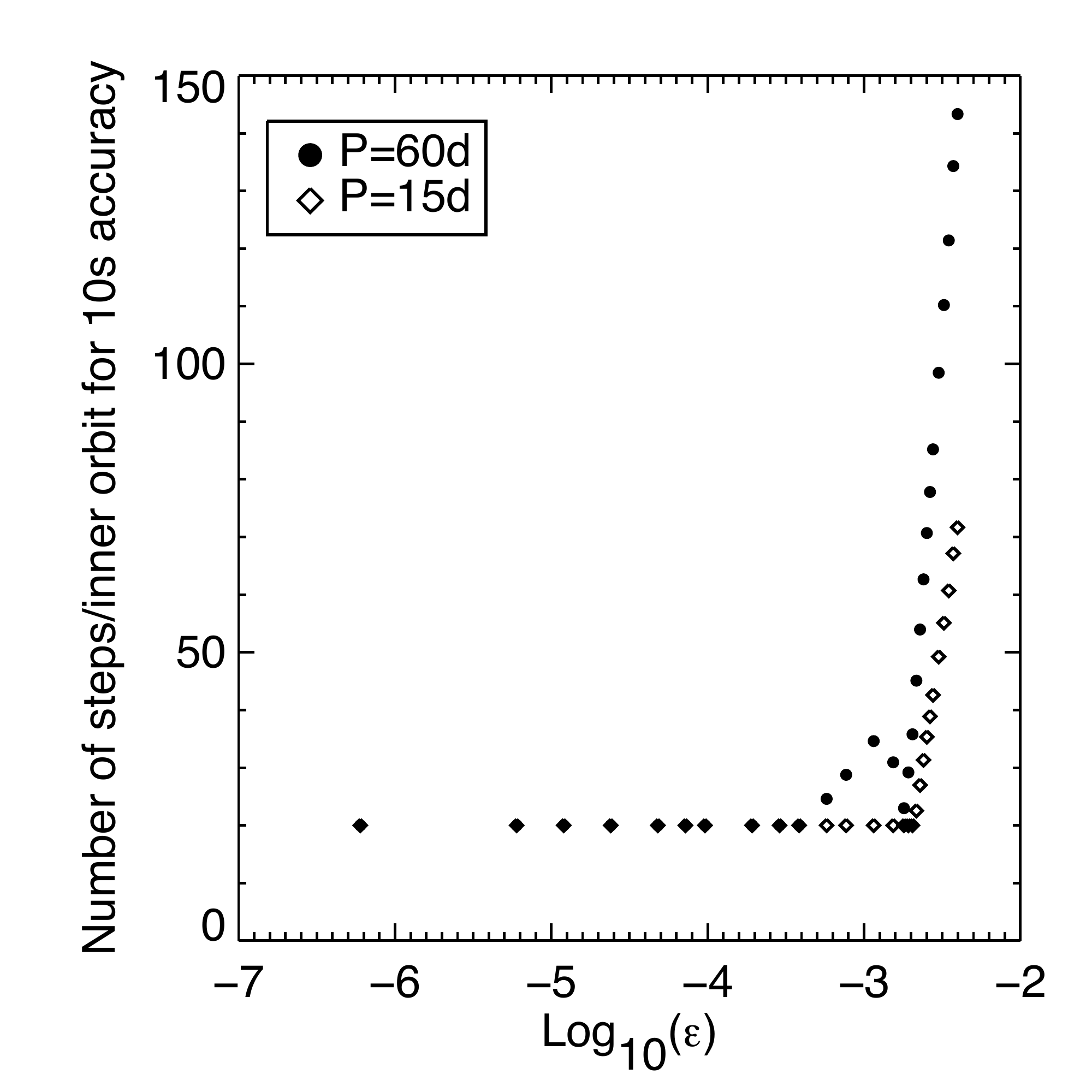}
\caption{Number of steps per orbit of our sample systems required to achieve $\lesssim$ 10s error on each of the transit times of the inner planet for an integration of $\sim200$ orbits. For very strongly perturbed systems with $P_{inner} = 60$d, approximately 145 steps per orbit are required to achieve this accuracy. The same system scaled to an orbital period of 15d requires $\approx 70 \sim 145 \sqrt{15/60}$ steps per orbit. The recommended minimum number of steps per orbit  is 20. }
\label{fig:scale}
\end{center}
\end{figure}

Using the scaling that the error using a time step of $\Delta t = P/N$ is proportional to $(P/N)^2$ and the measured errors for a system at some value of $N$, one can then infer approximately the number of steps per orbit required to reach arbitrary accuracy. The number of steps to reach errors less than 10s on each of the transit times is shown in Figure \ref{fig:scale} for a system with $P_{inner}$ = 15 days, and for $P_{inner} = 60$ days, using $\Delta t = P_{inner}/20$ as the minimum time step possible (for even smaller time steps the integrator is not always well-behaved, see the end of Section \ref{sec:inopt}.). 

\begin{figure}[ht]
\begin{center}
 \includegraphics[scale=0.4]{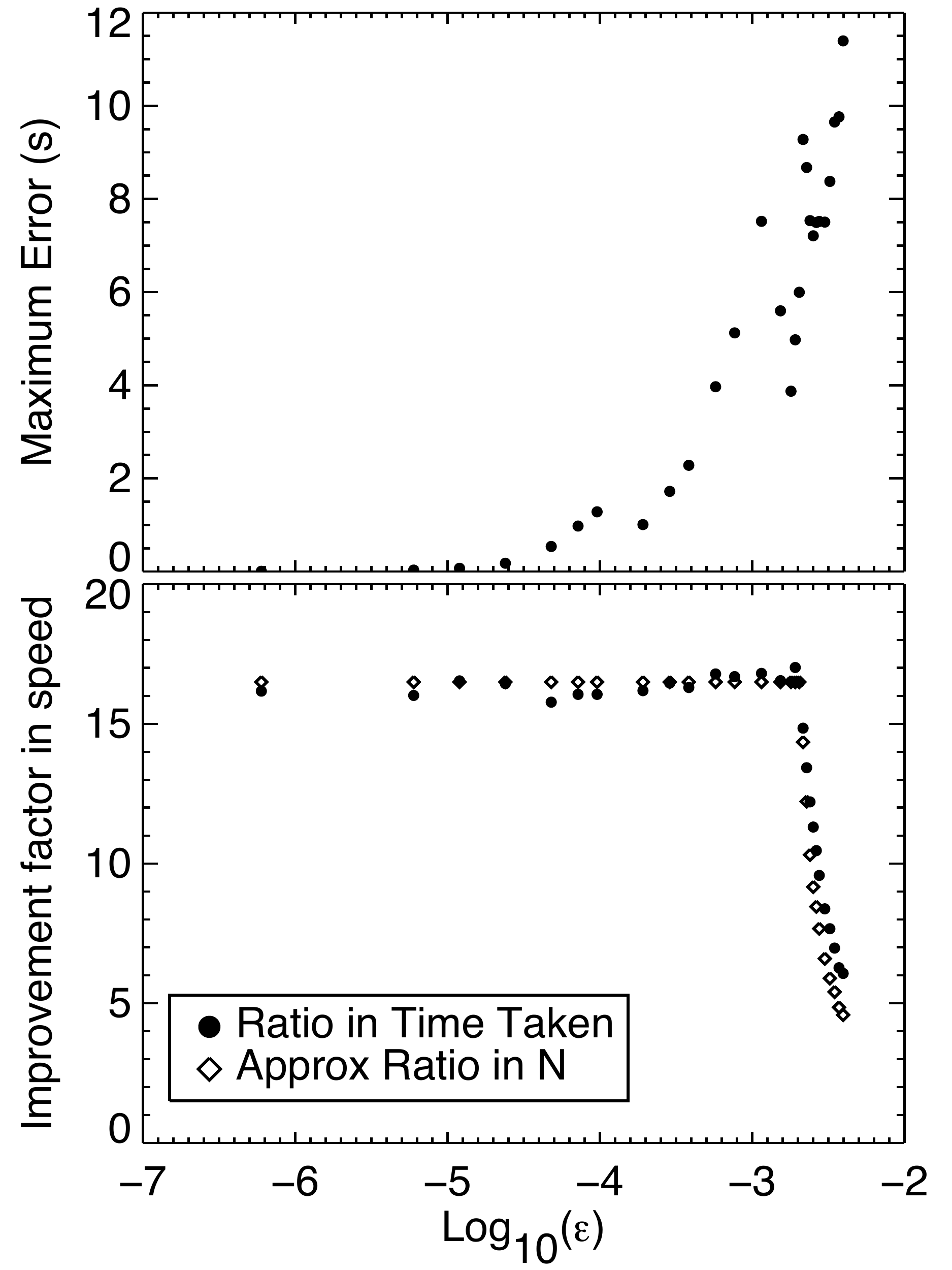}
\caption{Comparison of the computational time TTVFast to that of Bulirsch-Stoer. TTVFast is faster than our Burlisch-Stoer code by a factor of $\sim$5-20 for a wide variety of $\epsilon$. The upper panel shows the maximum error in a transit time over the integration (of 200 orbits, or $\sim$3000 days), as compared to the transit times measured by our Burlisch-Stoer code.  The lower panel shows the improvement factor in speed, or Time(Bulirsh-Stoer)/Time(TTVFast), using a time step as described in the text. The diamond symbols denote the predicted improvement if Bulirsch-Stoer effectively used 330 time steps per orbit.}
\label{fig:Scale2}
\end{center}
\end{figure}

Assuming the calculated transit times must be measured to within 10s, one can then use the time steps required to achieve this for different TTV amplitudes, and compare the computational time of TTVFast to Bulirsch-Stoer.  In other words, we integrate each of the systems (with $P_{inner} = 15$d) with TTVFast using the number of steps prescribed by the curve in open diamonds in Figure \ref{fig:scale}, and compare the computational time required to that of Bulirsch-Stoer for the same system. The results are shown in Figure \ref{fig:Scale2}. For a 15d period orbit, TTVFast is 10-20x faster than Bulirsch-Stoer for values of $\epsilon \lesssim 10^{-3}$ given the target of errors $\lesssim$ 10s. For longer orbital periods, the speed-up will be less significant, as more steps per orbit are required to reach $\lesssim$10s in error (as in Figure \ref{fig:scale}).

Since the computational time of fixed time step integrators scales linearly with the number of steps, we can say that our Bulirsch-Stoer code uses effectively $\sim16.5 \times$ more steps per orbit than TTVFast does in the best cases (when TTVFast uses only 20 steps per orbit). The speed-up predicted assuming that Bulirsch-Stoer uses 330 ``effective" steps per orbit is shown in Figure \ref{fig:Scale2} with the starred symbols. This is good approximation to the measured speed-up, although for the largest perturbations Bulirsch-Stoer requires slightly more ``effective" steps than 330. 

\section{Discussion}\label{sec:Discussion}
\subsection{The Kepler TTV sample}
\begin{figure}[[ht]h]
\begin{center}
 \includegraphics[scale=0.35]{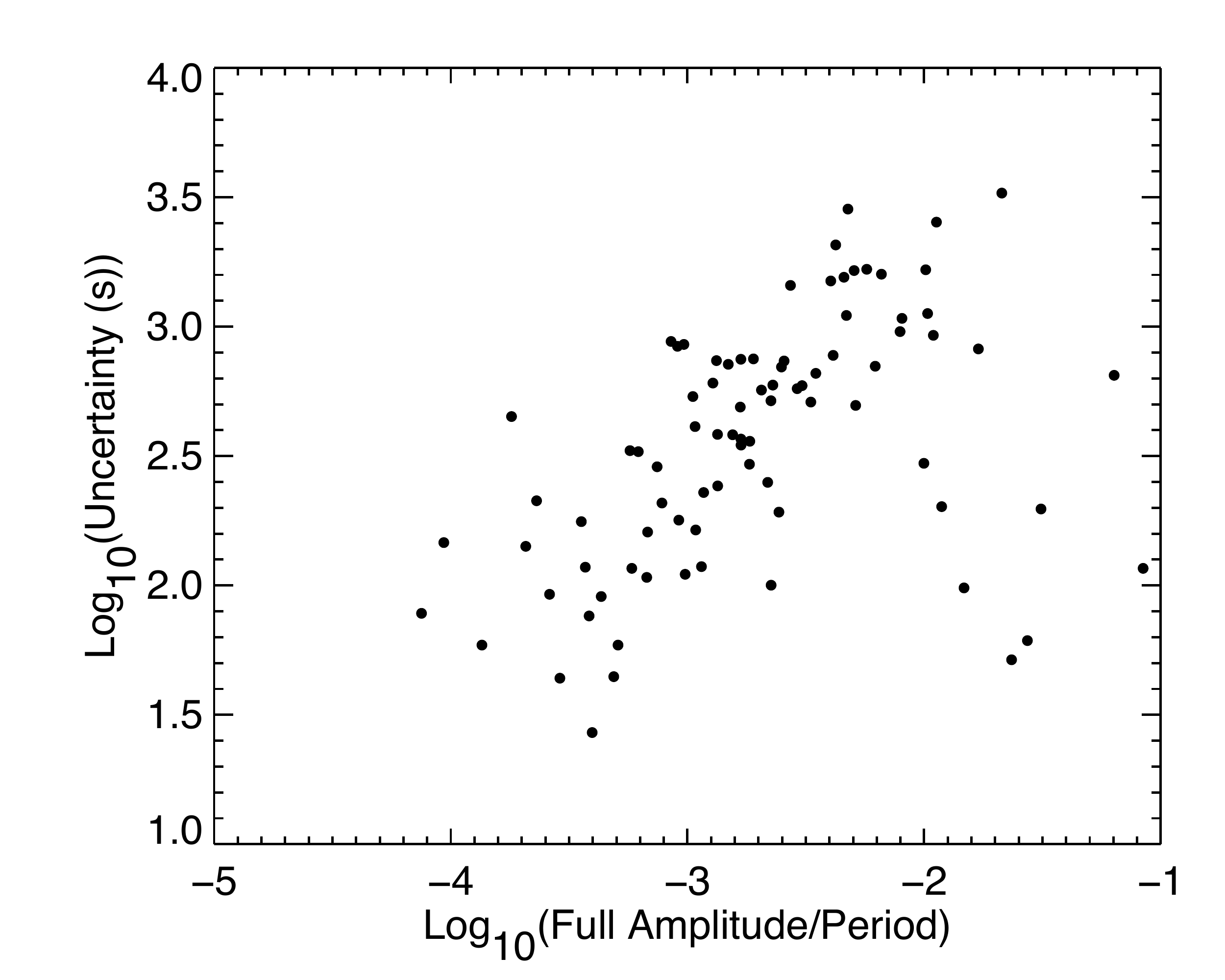}
\caption{The full amplitude of the TTVs, relative to the orbital period of the perturbed planet, and the transit time uncertainties for roughly 65\% of the Kepler planetary candidates with significant TTVs, or 85 systems in total, from \citet{Tsevi}. The remaining 35\% likely have TTVs with a period longer than the observational baseline and therefore have unconstrained amplitudes. Note that the TTV period scales with the orbital period. }
\label{fig:TTVs}
\end{center}
\end{figure}

 \citet{Tsevi} analyzed the first twelve quarters of the {\it Kepler} photometry data set (roughly 3/4 of the data) and compiled a list of the planet candidates with significant TTVs. Out of 130 that were flagged as significant, 85 of the TTV signals were sinusoidal in shape. Thirty-nine of the signals did not exhibit both a minimum and a maximum, suggesting that the entire period of the TTV had not been observed. These were fit with a parabola instead and do not have a constrained amplitude. Six systems were poorly fit by both a parabola and a sinusoid. Using their results, we show in Figure \ref{fig:TTVs} the range of full TTV amplitude, relative to orbital period, and uncertainty on the timing measurements for each of the 85 systems fit with a cosine function.
 
  The relative TTV amplitude is a proxy for $\epsilon$, but not a perfect one - other factors besides the mass and orbital separation affect the TTV amplitude. Given the typical sizes of the {\it Kepler} candidates,  the majority of these systems should have $\epsilon \lesssim 10^{-3}$. We note again that KOI-142b, which has the largest relative TTV observed, has a value of $\epsilon = 10^{-3.2}$.  For most of the {\it Kepler} targets, the timing uncertainties are several minutes. The median orbital period of these 85 targets is 14.6d, and 87\% have an orbital period less than 40 days. 
  
 These measurements indicate that TTVFast will perform well for all currently observed TTV systems, and guide our recommendations for the use of TTVFast. For most systems, using 20 steps per orbital period is sufficient given the uncertainties on the transit times, the sizes of the TTVs and the expected mass range of the planets, and the orbital periods. Using as few as 20 steps per orbit represents a significant increase in efficiency compared with Bulirsch-Stoer (as detailed in Section \ref{sec:BS}).  
 
 If higher accuracy on the transit times is needed, the user can employ a smaller time step.

\subsection{Analytic Methods based on Perturbation theory}\label{sec:analytic}
As mentioned in Section \ref{sec:Intro}, \citet{NesvornyMorbidelli}, \citet{N3} and \citet{NB10} have developed an analytic algorithm for determining transit times of nearly Keplerian orbits based on Hamiltonian perturbation theory.  Given a set of $(a,e,i)$, a set of Fourier coefficients is calculated. Any subsequent exploration in the angles (mean longitudes, pericenters, and nodes) and masses of the planets, as well as the incorporation of additional transits, is extremely fast as it just involves evaluation of the Fourier series. 

When searching parameter space for a solution, then, this method is significantly faster than $n$-body direct integration because given a set of $(a,e,i)$ one can search the angle dimensions essentially for free. This can be extremely useful when searching parameter space for the orbit of a non-transiting perturber. The perturbation method is also useful for understanding the source of various frequencies observed in TTVs, features harder to interpret with direct $n$-body integrations. Note that the perturbation theory method does not apply for systems very near resonances.

However, once a solution has been found, the characterization of the uncertainties in the parameters (often determined using MCMC) requires the full $n$-body solutions. In this case, a code like TTVFast will represent an improvement over other methods. 

\section{Conclusion}\label{sec:conclusion}
We have developed an efficient and accurate code for computing transit times of weakly perturbed Keplerian orbits around single stars. This code capitalizes on the fact that 1) symplectic integrators are significantly faster than most traditional integrators and that 2) transit times only need to be calculated to an accuracy small compared to the timing uncertainty. We leave it to the user to choose a time step which suits their need. 

In order to effectively invert a TTV signal, an entire cycle, or even more, must likely be observed. As the TTV period scales with the orbital period, most future TTV targets will also be systems with short period planets. The discussion and numerical explorations in Section \ref{sec:Numerical} indicate that for short period systems (those found by the {\it Kepler} telescope, and those which will be found by future transiting surveys), TTVFast can provide a significant increase in computational speed compared to Bulirsch-Stoer. 

 Our code is available in both C and Fortran at: {\bf http://github.com/kdeck/TTVFast}.

\acknowledgements{ We would like to acknowledge Dan Fabrycky for testing our code and the {\it Kepler} TTV team for helpful suggestions. K.M.D. acknowledges the support of an NSF Graduate Research Fellowship. E.A. acknowledges funding by NSF Career Grant AST 0645416, NASA Astrobiology Institute's Virtual Planetary Laboratory, supported by NASA under cooperative agreement NNH05ZDA001C,Ê and NASA Origins of Solar Systems Grant 12-OSS12-0011. Work by K.M.D and M.J.H. was supported by NASA under grant NNX09AB28G from the Kepler Participating Scientist Program and grants NNX09AB33G
and NNX13A124G under the Origins program. D.N. acknowledges support from NSF AST-1008890.}
\bibliographystyle{apj}
\bibliography{deck_ms_apj}

\clearpage

\end{document}